\documentclass[%
 reprint,
 amsmath,amssymb,
 aps,
]{revtex4-1}

\usepackage{graphicx}
\usepackage{dcolumn}
\usepackage{bm}
\usepackage{comment}    

\begin{document}

\preprint{APS/123-QED}

\title{Evolution of cooperation on scale-free networks under limited resources}

\author{Sara Sadat Veradi Isfahani}

\author{Farhad Fazileh}%
 \email{fazileh@cc.iut.ac.ir}
\affiliation{Department of Physics, Isfahan University of Technology, Isfahan 84156-83111, Iran}

\date{\today}

\begin{abstract}
Limitation of resources has been recently introduced as a mechanism for the survival and coexistence of cooperators with defectors in well-mixed populations. Here we examine the same model on a scale-free network. A prisoner's dilemma game on a scale-free network has shown coexistence of cooperators and defectors for the entire range of parameters. Our results show that by introducing the network to the limited resources model, the cooperator-dominated region in the parameter space expands comparing to the results of well-mixed population and the coexistence region becomes narrower. The effect of scale-free network is therefore interpreted as improving the cooperation in the population and reducing the coexistence.  
\end{abstract}

\pacs{02.50.Le,87.23.Kg}
\maketitle


\section{\label{introduction}Introduction}

Cooperative behavior is fundamental in evolution of complex systems from simpler units. The evolution of multicellular life from eukaryotic cells is an example. However, understanding the survival of cooperation in a world that selfish behavior normally provides higher benefits \cite{Darwin1859,Hamilton64a,Hamilton64b,Riolo01,Dawkins89} remains a challenge that has attracted a lot of attention during past two decades.  For this purpose evolutionary game dynamics \cite{Smith82} has been studied extensively for simple two-player games like the Prisoner's Dilemma (PD). In PD every individual selects one of two available strategies, cooperation or defection. Two cooperators in a play receive $R$, while two defectors receive $P$; and a cooperator confronting a defector receives $S$ and the defector in turn receives $T$, then $T>R>P>S$. It is clear that for this game it is better to defect, regardless of the opponent's strategy. According to the rules of the evolutionary game theory if each strategy is reproduced according to its payoff, whenever evolution under replicator dynamics \cite{Nowak06} takes place, we expect that the population of cooperators asymptotically vanishes in a well-mixed population (which every pair of individuals are equally likely to interact). 

In order to adopt our model to more realistic situations in which cooperators survive in the evolutionary process, different approaches have been proposed. Examples include departure from the well-mixed population regime or adoption of other games such as snowdrift game (SG), which are more favorable to cooperation. In SG the order of $P$ and $S$ is exchanged, such that $T>R>S>P$. Thus, contrary to the PD now the best action depends on the opponent's strategy. It is better to defect against a cooperator and cooperate against a defector.  Nowak and May are the pioneers of the scenario for departure from well-mixed population \cite{Nowak92}, where they included spatial structure in the PD, such that individuals are allowed only to interact with their immediate neighbors. Nowak has also introduced several different mechanisms for the survival of cooperation \cite{Nowak06}. One decade after the work of Nowak and May \cite{Nowak92} experiment confirmed that evolution of cooperation is affected by topological constraints in a sizable way \cite{Kerr02}. However, other studies \cite{Hauert04} using the SG has shown that contrary to PD, cooperation is inhibited when evolution in the SG is carried out on a spatial structure, which shows that spatial structure is not necessarily beneficial for the cooperative behavior. 

In a recent work \cite{Requejo12} another mechanism, the so-called limiting of resources, has been introduced as a mechanism for the survival and coexistence of cooperators with defectors. It was found that the limited resources in addition to imposing a finite population size, as usually is assumed in evolutionary game theory \cite{Taylor07,Axelrod81,Nowak98,Hauert05,Gomez07,Riolo01}, can also produce dynamic payoff matrices \cite{Requejo12,Lee11,Requejo11} and switch the payoff for the interactions between PD and Harmony Game.  

In this article we take another step and study the evolution of the limited resources model in which the payoff matrix for the interactions can switch between different games during the evolution of the population on a scale-free network to examine the effect of a network structure on the survival of cooperation in this model.

\begin{figure*}
\includegraphics[scale=0.5,angle=270]{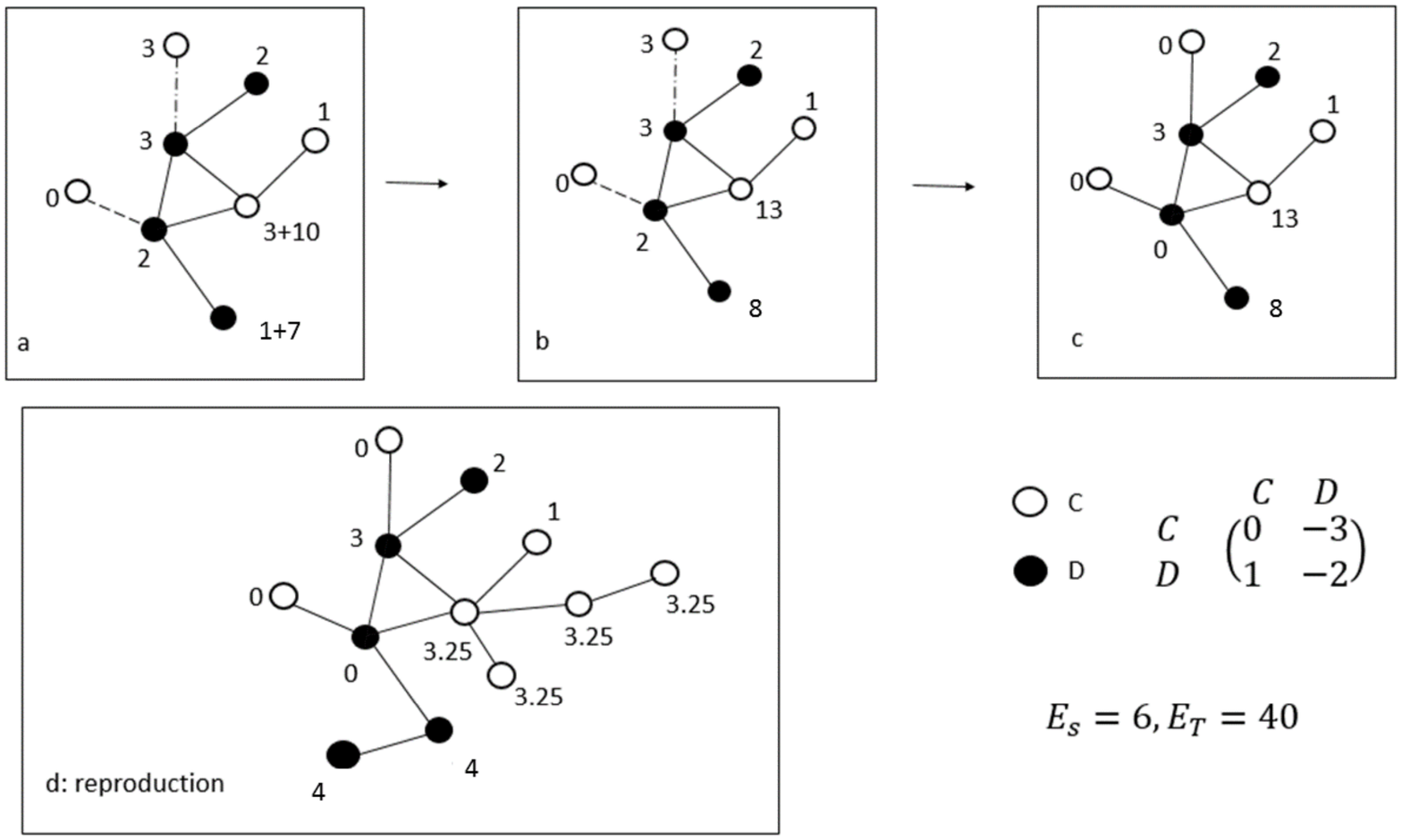}
\caption{\label{fig:scheme}This figure shows a simple example of limited resources algorithm on a network.
(a){\em Environment resource influx}: The environment provides resources
from the interval $[0,2E_T/N]$ to two randomly chosen nodes, regardless of
their strategy or the amount of resource they already have. (b) {\em Playing}: This step
shows two interactions following the rules stated and the payoff matrix shown in the
picture. The dashed-doted line shows an interaction between a cooperator and a defector.
According to the matrix the cooperator looses 3 units of it's resources and the defector 
gains a total amount of $\Delta E$. The dashed line also shows an interaction between  
a cooperator and a defector. Here the defector has the required resources for entering 
the game, but the cooperator does not have any resources left after the play. This means that the defector pays
the cost but does not gain any rewards, so the total payoff for the defector in this 
game will be -2 reducing it's resources to zero. (d) {\em Reproduction}: In this step every individual
possessing a resource more than $E_s=6$ will reproduce an individual with the same strategy.
The amount of resources belong to the two of these individuals after one reproduction process 
will be half of the mother's resource before reproduction. In this example reproduction
process can occur for a cooperator
and the defector. The cooperator has a resource equal to 11, so only one new player will be
added to the network. The defector has resource of 13, so two steps of reproduction 
is required. }
\end{figure*}

\section{\label{model}The model}
In our model, initially a scale-free network is constructed using the following \textit{growth} and \textit{preferential attachment} rules, associated with the Barabasi and Albert (BA) model \cite{Barabasi99}. We start from a small number of nodes $(m_0)$ and at each time step we add a new node with $m \leq m_0$ links from the new node to $m$ different nodes already present in the network (\textit{growth}). For choosing the nodes which the new node connects to, we assume that the probability $p_i$ that a new node will be connected to node $i$ is $p_i=k_i/\sum k_i$, where $k_i$ is the degree of node $i$ (\textit{preferential attachment}). 

In contrast to other studies that a Moran process is normally applied to a fixed network, in our model the network is dynamic during the simulation which means that nodes are added and removed from the network during the evolution of the system. This evolution of the network is necessary in modeling the mechanism of limited resources. When the network is constructed, individuals with two strategies, namely cooperative (C) and defective (D) are placed randomly on the nodes of this network. Each individual is given a number indicating its internal resources. In each time step neighboring individuals interact and exchange resources with each other according to the following rules. The rules are similar to those introduced in reference \cite{Requejo12}.  
\subsection{Update rules}
The updating consists of the following steps:
\paragraph{Environment resource influx}
The environment provides resources in the average portions of $E_T/N$ in each time step to two randomly selected individuals from the population independent of their strategy; where $E_T$ is a constant total resource influx and $N$ is the number of individuals. Specifically, in our simulation the resources provided by environment is randomly selected from the interval $[0,2E_T/N]$ (to prevent sudden explosion of population after $E_s/(E_T/N)$ time steps, where $E_s$ is the production threshold and it will be introduced latter).
\paragraph{Playing}
Every two neighboring individuals on the network interact in each time step. Defectors spend a cost $E_c$ to enter the game and steal a reward of $E_r$ from the co-player. If a defector's internal resource is less than $E_c$, the player does not pay the cost nor receive the reward. If the defector's co-player has resources less than the reward, the entire amount of co-player's internal resources will transfer to the defector. So the pay-off matrix for the interaction is:
\begin{equation}
\label{CD}
\begin{matrix}
  & C \qquad D    \\
\begin{matrix} C  \\  \\  D  \end{matrix}  &  
\begin{pmatrix}
0 & & -E_r \\   \\
\Delta E & &  -E_c
\end{pmatrix} 
\end{matrix}
\end{equation}
where $\Delta E = E_r - E_c>0$ is the net reward for defectors in interactions with cooperators (payoffs for the row player). So, as long as the C player internal resources are more than $E_r$, the interaction equals a simplified PD game.
\paragraph{Reproduction}
In each time step individuals who have internal resources more than a threshold value, $E_s$, will reproduce an individual with the same strategy and half of their internal resources will transfer to the new born individual. The new born individual is connected to her mother and $m-1$ randomly selected other nodes with a probability proportional to the degree of the nodes (degree of a node is the number of links connected to that node). 
\paragraph{Death}
In each time step two individuals are selected randomly and are removed from the network with a small probability of $f$. 

The above rules are summarised in Fig.~\ref{fig:scheme}.

\begin{figure}[b]
\includegraphics[scale=0.3,angle=270]{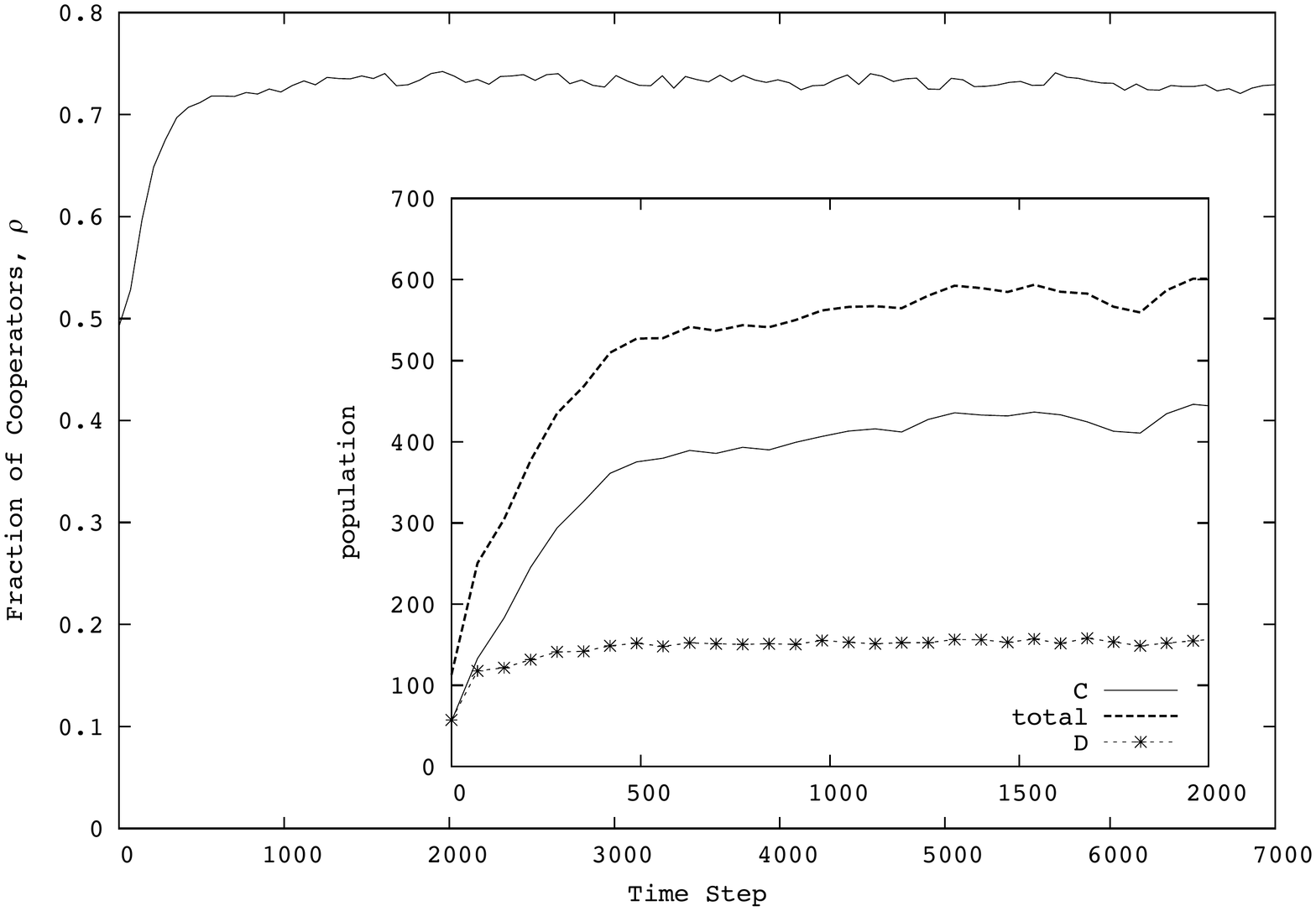}
\caption{\label{fig:a2a_rho} Fraction of cooperators, $\rho_C$, versus time for a population on an all-to-all network interacting through limited resources model. In this graph the reproduction threshold is 1000, initial population is 100, environment resource influx is 20000, death frequency is 0.01, the cost paid by defectors or $E_C$ is 400, $\Delta E = 200$, and the initial fraction of cooperators is 0.5. Here we have recorded the results once after 70 time steps.}
\end{figure}

\begin{figure}[b]
\includegraphics[scale=0.3]{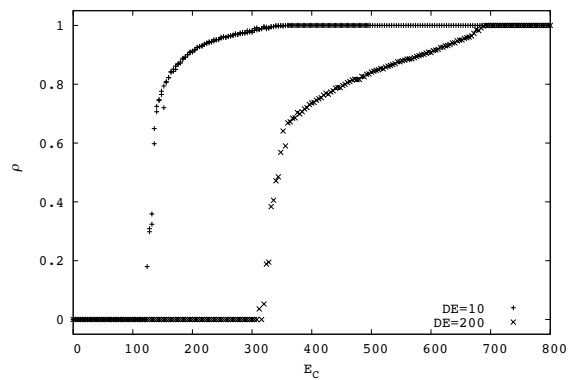}
\caption{\label{fig:a2a_del_e} Final fraction of cooperators, $\rho_C$, after playing in a well-mixed population as a function of the cost paid by defectors, $E_C$, for two values of $\Delta E=10$ and $\Delta E=200$ . In both graphs the reproduction threshold is 1000, initial population is 100, environment resource influx is 20000 and the initial fraction of cooperators is 0.5. }
\end{figure}

\begin{figure}[t]
\includegraphics[scale=0.3]{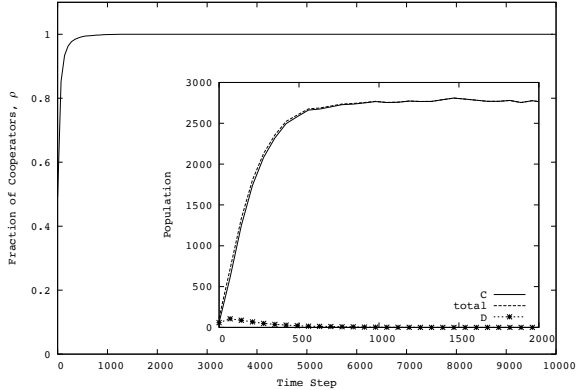}
\caption{\label{fig:sf_rho} The main graph shows fraction of cooperators, $\rho_C$, versus time for playing on a scale-free graph. Inset shows the population of cooperators(line), defectors(dashed-plus), and the total population(dashed line) versus time. Here the reproduction threshold is 1000, initial population is 100, environment resource influx is 20000, death frequency is 0.01, the cost paid by defectors or $E_C$ is 400 ,the initial fraction of cooperators is 0.5 and $\Delta E = 200$. Here we recorded the results once after 70 time steps.}
\end{figure}

\begin{figure}[b]
\includegraphics[scale=0.3]{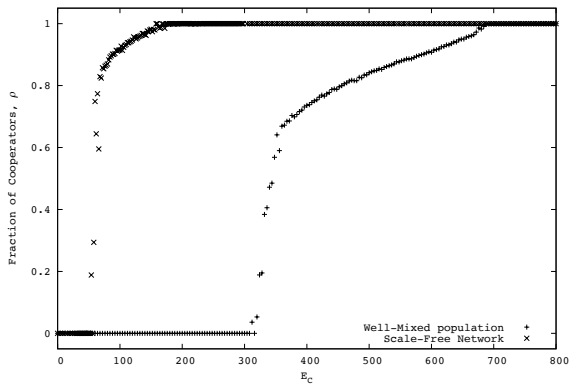}
\caption{\label{fig:sf_wm} Final fraction of cooperators, $\rho_C$, as a function of $E_C$ for playing on a scale-free network and in a well-mixed population. In these graphs the reproduction threshold is 1000, initial population is 100, environment resource influx is 20000, death frequency is 0.01, the initial fraction of cooperators is 0.5 and $\Delta E=200$. Results are averaged over 10 independent runs of the program.}
\end{figure}

\section{Simulation results}

In order to simulate the evolution of the system we start with a population of N players with $\rho$ indicating the fraction of cooperators or C players. According to the model introduced in ref.~\cite{Requejo12}, and since the number of interactions in each time step is supposed to be proportional to the number of individuals, it was assumed that there are $N/2$ interactions in each time step. Each interaction consists of four processes with the following order. First, two random individuals are selected for consuming environment resource influx, then two nearest neighbor individuals are selected randomly for a play, then reproduction and death processes are taking place as required. The evolution of the population and its network is continued until a stable population with reasonable fluctuations is reached (in our calculations that starts with a population of 100 individuals, total environment resource influx of 20000, death rate of 0.01, and reproduction threshold of 1000, this condition is normally reached after around 2000 time steps).  In order to compare our model with similar calculations from other authors, we establish an all-to-all network where each player is connected to all other existing players with a probability of one. Interactions on this network is expected to be identical with interactions in a well-mixed population. This means that for specific parameter values we expect to have coexistence of C and D players in a final meta-stable state \cite{Requejo12}. When all parameters are fixed, increasing $\Delta E$ will decrease the final value of $\rho_C$. These results are consistent with the data shown in Fig.~2 of ref.~\cite{Requejo12}.

Summary of our results  are shown in Fig.~\ref{fig:a2a_rho} and Fig.~\ref{fig:a2a_del_e}. These figures show the results of our program for an all-to-all network. The results are persuasively in agreement with Requejo et. al. \cite{Requejo12}.

It is shown that the limited resources model on a well-mixed population provides different regions of coexistence for different values of parameters. In fact, provided the meta-stable state is reached and the only variable is $\Delta E$, it is shown in Fig.~\ref{fig:a2a_del_e} that by decreasing $\Delta E$ this region shrink, and it is shifted toward the smaller values of $E_c$, in agreement with  ref.~\cite{Requejo12}. 

To simulate limited resources on a scale-free network, we need to consider the dynamic features of the model. That means, reproduction of a new player adds a new node to the network. This node is connected to $m$ existing nodes or players. Death also removes a node and all its connections from the network. Considering these points and using the method described above, we observed that by adding such a dynamic network to our model, meta-stable states are reached sooner; that is, with identical parameter values meta-stable states of limited resource model on a scale-free network results with less time steps in comparison with well-mixed population. The described results are shown in Fig.~\ref{fig:sf_rho}.

Another observation is related to the parameter values that allow coexistence in meta-stable states. Considering identical parameter values, we can see that playing on a scale-free network causes the population to establish cooperation at lower values of $E_C$ as is shown in Fig.~\ref{fig:sf_wm}. Also we can see that the region of coexistence shrinks when we add the scale-free network to the model. In other words limited resources on a scale-free network helps cooperation to establish and dominate faster.

\section{ Discussion}
To summarise the results we should mention that limited resources helps cooperation to stay in the population. The final fraction of cooperators depends on the parameter values. Environment resources effectively controls the reproduction rate. The effective rate of death is controlled by death frequency. In a well-mixed population any two individuals can be the players of the game; however, it seems more logical if individuals that are related in some way be allowed to play with each other. Limited resources model is associated with a dynamic population. To ease the simulation of this feature and in order to choose a network that describes many natural phenomena, a scale-free network has been chosen and added to the limited resources model. 

Generally a network helps cooperation to evolve and stay in the population. This feature has been tested with different mechanisms. In fact a network causes cooperators to reach a kind of advantage by gathering and forming clusters ~\cite{Nowak06,Santos05}. In a scale-free network, cooperators occupying hobs are able to promote cooperation by making clusters of cooperators around hobs~\cite{Santos05}. In such a situation, neighbors are mainly relatives. In this model relatives which have the same strategies are produced and connected to each other through the reproduction process. Also, as the time passes, new individuals are replaced in the population. This means cooperators placed on hobs can effectively interact with their relatives. These cooperators may interact with defectors too, but the passage of time guarantees that cooperators can achieve a kind of advantage associated with the resources, because eventually the average resources of the defectors decreases, which causes the number of defectors allowed to play decrease too.

Simulations show that limited resources algorithm combined with scale-free network is beneficial for evolution of cooperation. 
In fact comparing a well-mixed population interacting through limited resources algorithm with a population on a scale-free network interacting through the same algorithm shows an increase in the final frequency of cooperators. 
Fig.~\ref{fig:sf_wm} shows that cooperation starts to resist defection, and also stay in the population at lower values of $E_C$. Also the meta-stable state region shrinks by introducing the network to the population.
At the end we may notice that knowing the exact mechanism of evolving cooperation in this model needs further studies. To find clusters of cooperators, the network should be explored. Also, it should be mentioned that the above mentioned network remains a scale-free network during the evolution.

The authors would like to thank Dr. K. Aghababaei Samani for useful discussions. F. Fazileh acknowledges financial support from Iran's National Elites Foundation.

\nocite{*}

\bibliography{our_VF_article}

\end{document}